\documentclass{article}
\usepackage{graphicx}
\usepackage[english]{babel}

\title{
\includegraphics[width=0.35\textwidth]{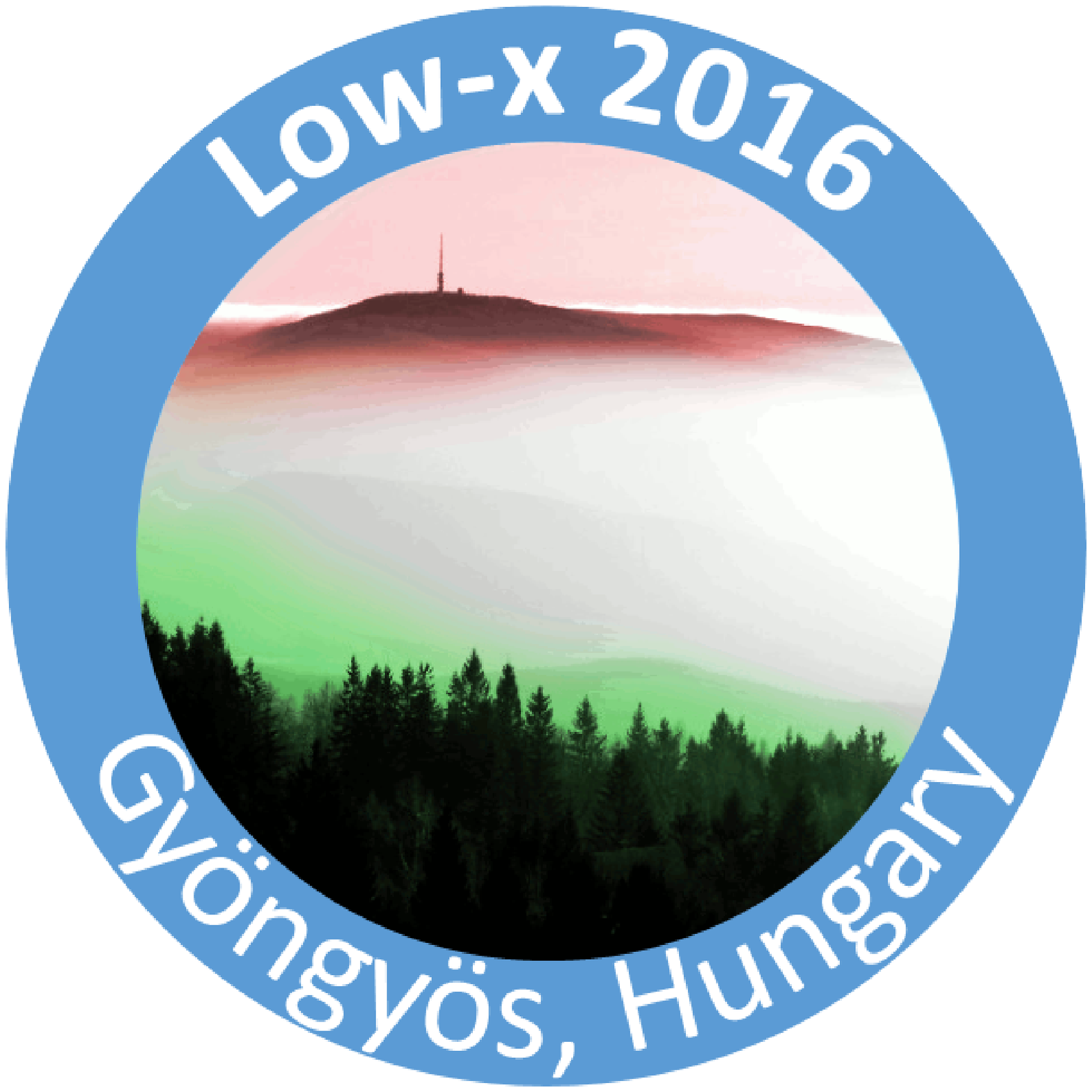}\\[1cm]
Double-parton scattering effects in double charm production within gluon fragmentation scenario}
\author{{Rafa{\l} Maciu{\l}a$^1$, Antoni Szczurek$^{1,2}$,}\\[1ex]
$^1$Institute of Nuclear Physics PAN, PL-31-342 Krak{\'o}w, Poland\\
$^2$University of Rzesz\'ow, PL-35-959 Rzesz\'ow, Poland\\
}

\begin{document}

\fontfamily{lmss}\selectfont
\maketitle

\begin{abstract}
  We discuss charm $D^0 D^0$ meson-meson pair production in the forward rapidity region
related to the LHCb experimental studies at $\sqrt{s}$ = 7 TeV.
We consider double-parton scattering mechanisms of double $c \bar
c$ production and subsequent standard $cc \to D^{0}D^{0}$ scale-independent hadronization as well as new double $g$ and mixed
$g c\bar c $ production mechanisms with $gg \to D^{0}D^{0}$ and $gc \to D^{0}D^{0}$ scale-dependent hadronization.
The new scenario with gluon fragmentation components results also in a new single-parton scattering mechanism of $gg$ production which is also taken here into account. Results of the numerical calculations are compared with the LHCb data for several correlation observables.
The new mechanisms lead to a larger cross sections and to slightly different shapes of the calculated correlation observables.
\end{abstract}

\section{Introduction}

Some time ago we have predicted that at large energies, relevant for the LHC, production of double charm should be dominated
by the double-parton scattering (DPS) mechanism \cite{Luszczak:2011zp}.
Afterwards, those leading-order (LO) collinear predictions were extended
to the $k_t$-factorization approach that effectively includes
higher-order QCD effects \cite{Maciula:2013kd,vanHameren:2014ava}.
The improved studies provide a relatively good description of the LHCb experimental data \cite{Aaij:2012dz}.
Besides, the single-parton scattering (SPS) $g g \to c \bar c c \bar c$ mechanism was found to be much smaller than the DPS one,
and is not able to explain the LHCb double charm data \cite{vanHameren:2014ava,vanHameren:2015wva}.

The theoretical analyses introduced above were based on the standard $c \to D$ hadronization scenario with scale-independent Peterson fragmentation function (FF) \cite{Peterson:1982ak}.
An alternative approach for hadronization effects is to apply scale-dependent FFs of a parton (gluon, $u,d,s,\bar u, \bar d, \bar s, c, \bar c$) to $D$ mesons proposed by Kniehl et al. \cite{Kniehl:2005de,Kniehl:2006mw}, that undergo DGLAP evolution equations. Both prescriptions were found to provide a very good description of the LHC data on inclusive 
$D$ meson production at not too small transverse momenta (see e.g. Refs.~\cite{Maciula:2013wg,Nefedov:2014qea}). In the latter approach, a dominant contribution comes from $g \to D$ fragmentation that appears in the evolution of the scale-dependent FFs and the $c \to D$ component is damped with respect to the scale-independent fragmentation scheme.   

The presence of the gluonic components modify the overall picture for the double charm production.
In the (new) scenario with scale-dependent hadronization the number of contributing DPS processes
grows. In addition, a new single-parton scattering mechanism SPS $gg \to DD$ appears.
Taking into account gluon fragmentation components there are more processes for single $D$ meson
production (two dominant components $g,c \to D$) and as a consequence many more processes for DPS
$DD$ production appear. Now there are three classes of DPS contributions.
In addition to the coventional DPS $cc \to DD$, discussed very carefully in Refs.~\cite{Maciula:2013kd,vanHameren:2014ava,vanHameren:2015wva}
there is a double $g \to D$ fragmentation mechanism, called here
DPS $gg \to DD$ as well as the mixed DPS $gc \to DD$ contribution.

Here the gluon and digluon production is
considered in the $k_t$-factorization approach with reggeized gluons
in the t-channel \cite{Nefedov:2013ywa} via subprocesses $RR\to g$
and $RR\to g g$, where $R$ is the reggeized gluon. We use scale-dependent fragmentation functions of Kneesch-Kniehl-Kramer-Schienbein (KKKS08)
\cite{Kneesch:2007ey} as implemented in the code available on the Web \cite{webpage}.
All details of the calculations presented here can be found in our original paper \cite{Maciula:2016wci}.

%-----------------------------------------------------------------
\section{A sketch of the theoretical formalism}
%-----------------------------------------------------------------

%-----------------------------------------------------------------------------
\begin{figure}[!h]
\begin{center}
\begin{minipage}{0.22\textwidth}
 \centerline{\includegraphics[width=1.0\textwidth]{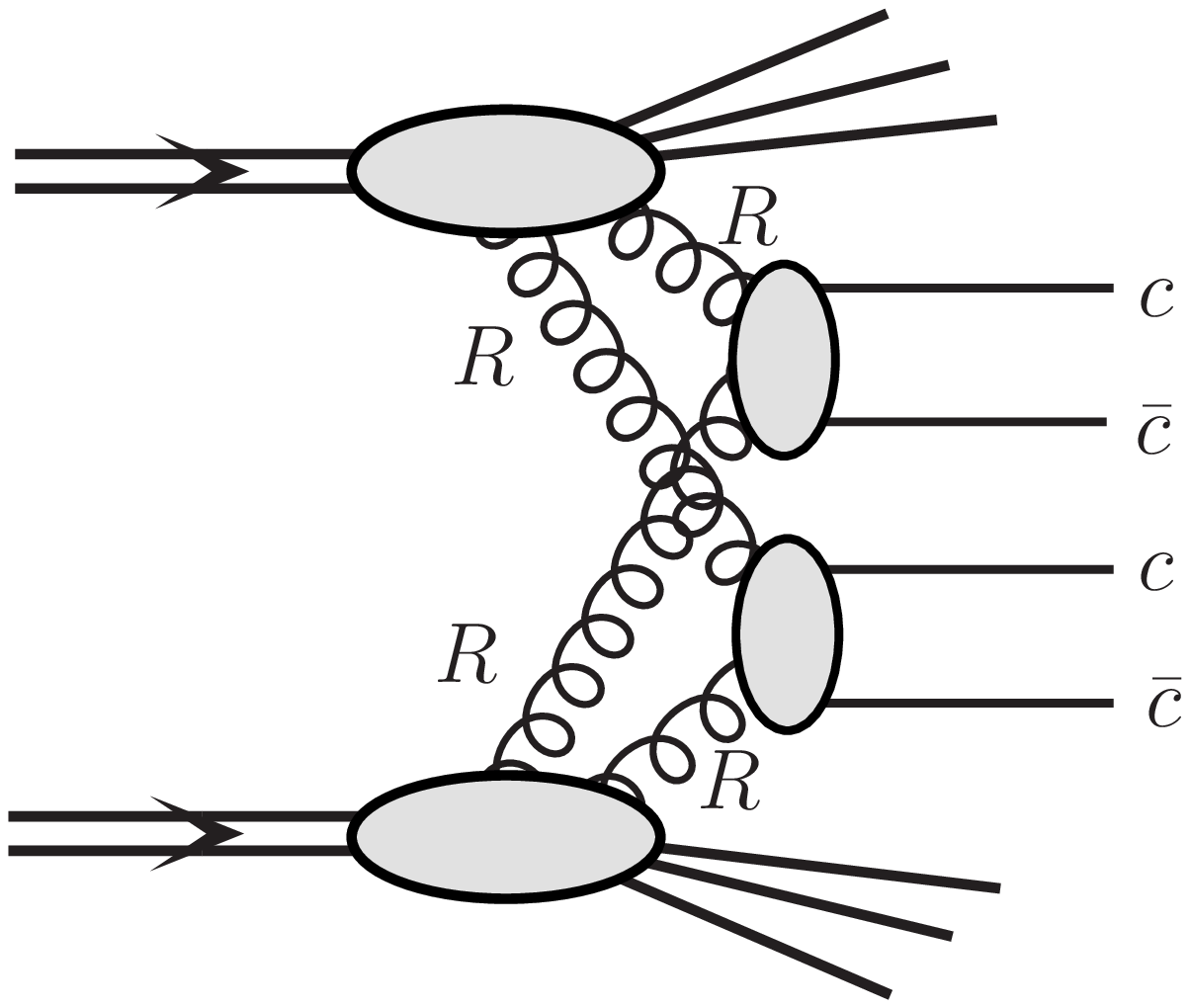}}
\end{minipage}
\hspace{0.0cm}
\begin{minipage}{0.22\textwidth}
 \centerline{\includegraphics[width=1.0\textwidth]{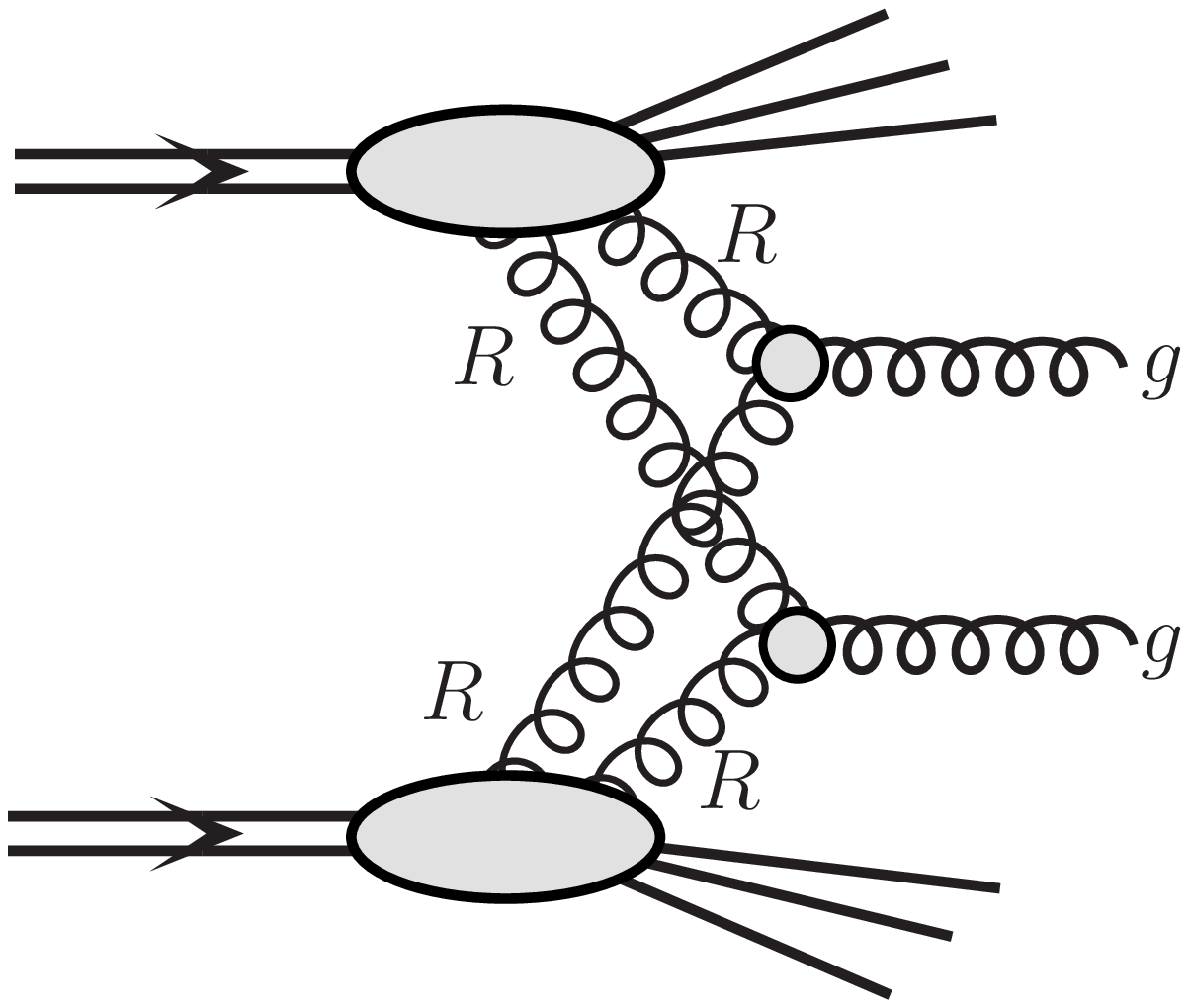}}
\end{minipage}
\vspace{0.0cm}
\begin{minipage}{0.22\textwidth}
 \centerline{\includegraphics[width=1.0\textwidth]{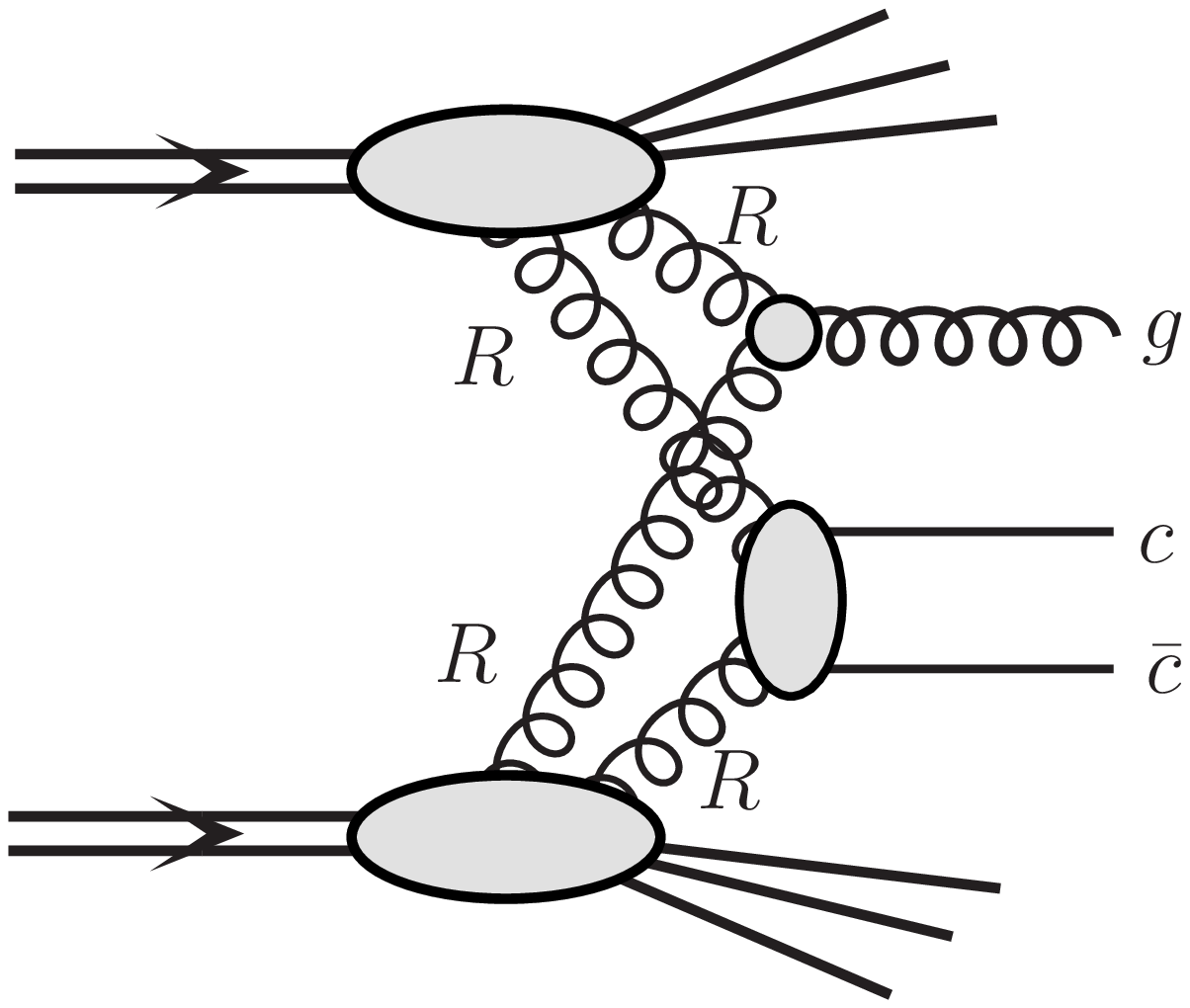}}
\end{minipage}
\hspace{0.0cm}
\begin{minipage}{0.22\textwidth}
 \centerline{\includegraphics[width=1.0\textwidth]{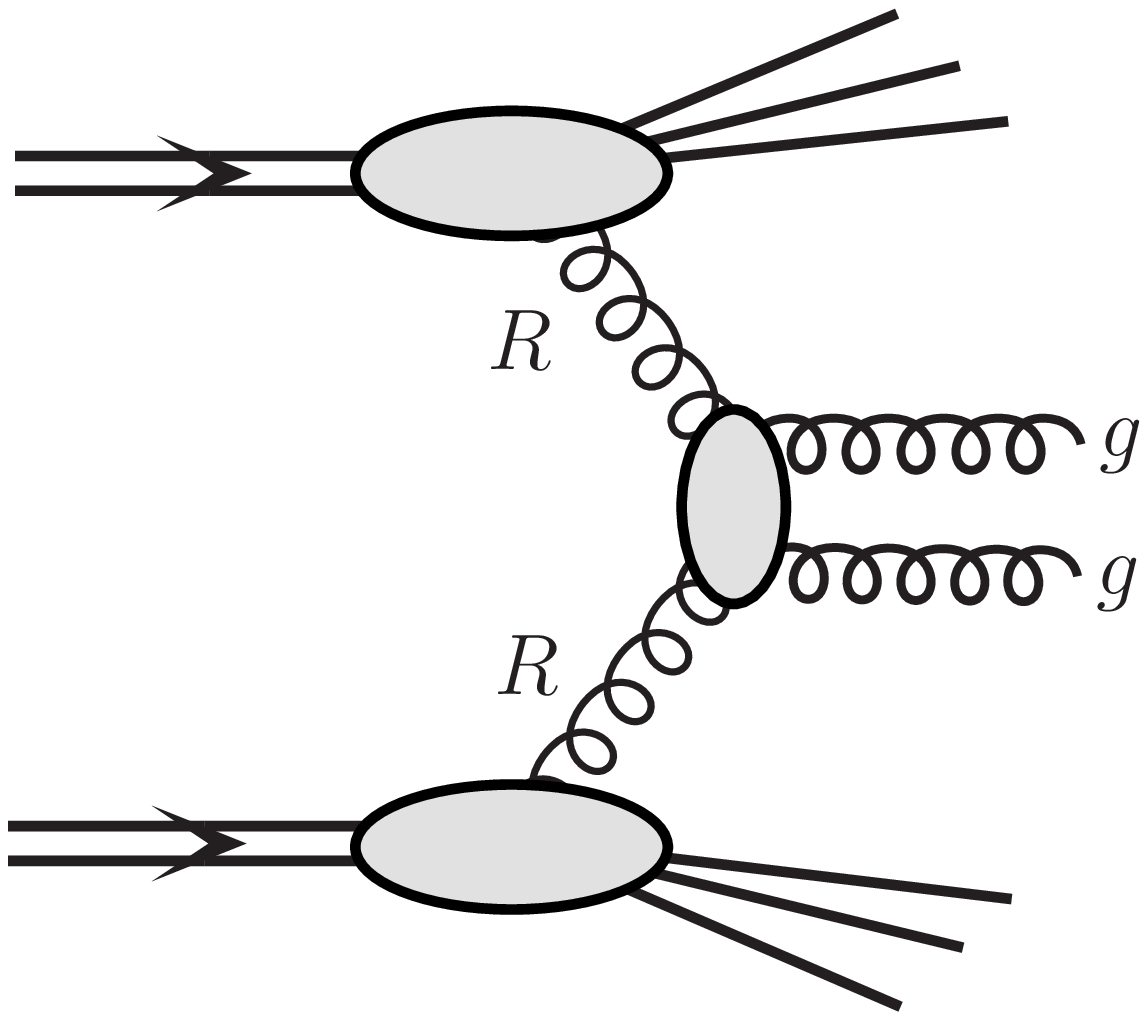}}
\end{minipage}
   \caption{
\small A diagrammatic illustration of the considered mechanisms.
 }
 \label{fig:diagrams}
 \end{center}
\end{figure}
%------------------------------------------------------------------------------

We will compare numerical results for $D^{0}D^{0}$ meson-meson production obtained with the two different fragmentation scenarios.
According to the scheme with scale-dependent FFs more processes for single $D$ meson production ($c$ and $g \to D$ components) has to be taken into consideration.
This also causes an extension of the standard DPS $DD$ pair production by new mechanisms.
In addition to the coventional DPS $cc \to DD$ (left diagram in Fig.\ref{fig:diagrams})
considered in Refs.~\cite{Maciula:2013kd,vanHameren:2014ava,vanHameren:2015wva}
there is a double $g \to D$ (or double $g \to \bar D$) fragmentation mechanism, called here
DPS $gg \to DD$ (middle-left diagram in Fig.\ref{fig:diagrams}) as well as
the mixed DPS $gc \to DD$ contribution (middle-right diagram in Fig.\ref{fig:diagrams}).

As a consequence of the new approach to fragmentation a new SPS $gg \to DD$ mechanism shows up (right diagram in Fig.\ref{fig:diagrams}). In this case the two produced gluons are correlated in azimuth and the mechanism will naturally lead to an azimuthal correlation between two $D$ mesons.
Such a correlation was actually observed in the LHCb experimental data \cite{Aaij:2012dz} and so far could not be explained theoretically.

DPS cross section for production of $cc$, $gg$ or $gc$ system, assuming factorization of the
DPS model, can be written as:
\begin{eqnarray}
\frac{d \sigma^{DPS}(p p \to c  c  X)}{d y_1 d y_2 d^2
p_{1,t} d^2 p_{2,t}} = 
\frac{1}{2 \sigma_{eff}} \cdot \frac{d \sigma^{SPS}(p p \to c \bar c
X_1)}{d y_1  d^2 p_{1,t}} \cdot \frac{d \sigma^{SPS}(p p \to c \bar
c X_2)}{d y_2 d^2 p_{2,t}}, \label{DPScc_factorization_formula}
\end{eqnarray}
\begin{eqnarray}
\frac{d \sigma^{DPS}(p p \to g g  X)}{d y_1 d y_2 d^2
p_{1,t} d^2 p_{2,t}} =
\frac{1}{2 \sigma_{eff}} \cdot \frac{d \sigma^{SPS}(p p \to g
X_1)}{d y_1  d^2 p_{1,t}} \cdot \frac{d \sigma^{SPS}(p p \to g
X_2)}{d y_2 d^2 p_{2,t}}. \label{DPSgg_factorization_formula}
\end{eqnarray}
\begin{eqnarray}
\frac{d \sigma^{DPS}(p p \to g c  X)}{d y_1 d y_2 d^2
p_{1,t} d^2 p_{2,t}} =
\frac{1}{\sigma_{eff}} \cdot \frac{d \sigma^{SPS}(p p \to g
X_1)}{d y_1  d^2 p_{1,t}} \cdot \frac{d \sigma^{SPS}(p p \to c \bar
cX_2)}{d y_2 d^2 p_{2,t}}. \label{DPSgg_factorization_formula}
\end{eqnarray}

The often called pocket-formula is a priori a severe approximation. The flavour, spin and color correlations may lead, in principle, to interference effects that result in its violation as discussed e.g. in Ref.~\cite{Diehl:2011yj}. Even for unpolarized proton beams, the spin polarization of the two partons from one hadron
can be mutually correlated, especially when the partons are relatively close in phase space (having comparable $x$'s). Moreover, in contrast to the standard single PDFs, the two-parton distributions have a nontrivial color structure which also may lead to a non-negligible correlations effects. 
Such effects are usually not included in phenomenological analyses. They were exceptionally discussed in the context of double charm production \cite{Echevarria:2015ufa}.
However, the effect on e.g. azimuthal correlations between charmed quarks was found there to be very small, much smaller than effects of the SPS
contribution associated with double gluon fragmentation discussed here.
In addition, including perturbative parton splitting mechanism also leads to a breaking of the pocket-formula \cite{Gaunt:2014rua}.
This formalism was so far formulated for the collinear leading-order
approach which for charm (double charm) may be a bit academic as it
leads to underestimation of the cross section.
Imposing sum rules also leads to a breaking of the factorized Ansatz
but the effect almost vanishes for small longitudinal momentum fractions 
\cite{Golec-Biernat:2015aza}. Taken the above arguments we will use the pocket-formula in the
following.

All the considered mechanisms (see Fig.~\ref{fig:diagrams}) are calculated in the $k_t$-factorization approach with off-shell initial state partons and unintegrated ($k_{t}$-dependent) PDFs (unPDFs). Fully gauge invariant treatment of the initial-state off-shell
gluons and quarks can be achieved in the $k_t$-factorization approach only when
they are considered as Reggeized gluons or Reggeons. We use the LO Kimber-Martin-Ryskin (KMR) unPDFs, generated from the LO set of a up-to-date MMHT2014 collinear PDFs fitted also to the LHC data (for more details see Ref.\cite{Maciula:2016wci}). 

%--------------------------------------------------
\section{Comparison to the LHCb data}
\label{results}
%--------------------------------------------------

\begin{figure}[!h]
\begin{minipage}{0.47\textwidth}
 \centerline{\includegraphics[width=1.0\textwidth]{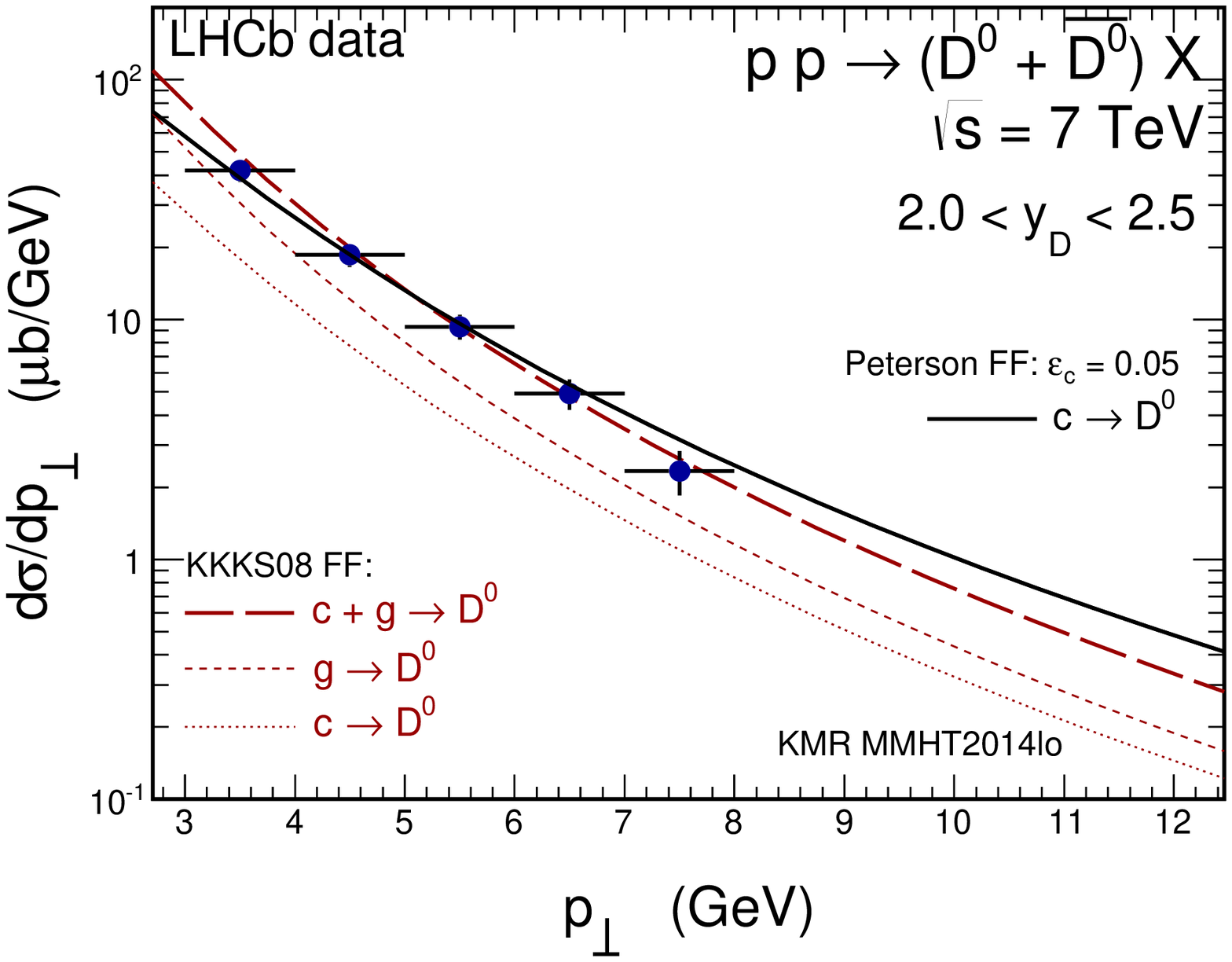}}
\end{minipage}
\hspace{0.5cm}
\begin{minipage}{0.47\textwidth}
 \centerline{\includegraphics[width=1.0\textwidth]{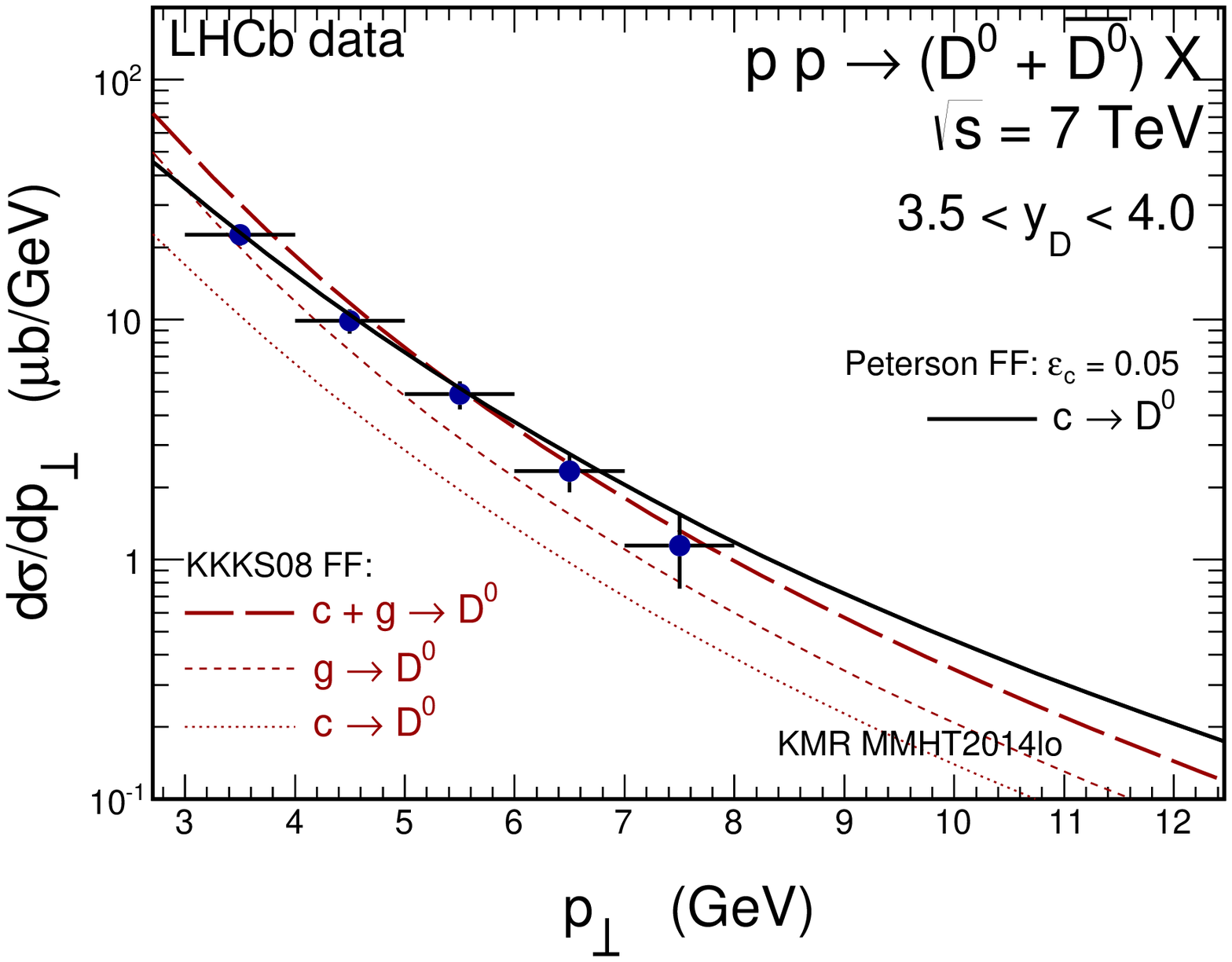}}
\end{minipage}
   \caption{
\small Charm meson transverse momentum distribution within the LHCb acceptance for inclusive single $D^{0}$ mesons (plus their conjugates) production.
The left and right panels correspond to two different rapidity intervals. 
Theoretical predictions for the Peterson $c \to D$ fragmentation function (solid lines)
are compared to the second scenario calculations with the KKKS08 fragmentation functions (long-dashed lines)
with $c \to D$ (dotted) and $g \to D$ (short-dashed) components
that undergo DGLAP evolution equation.
 }
 \label{fig:pTsingle}
\end{figure}

We start this section with a revision of inclusive single $D^{0}$ meson production measured some time ago by the LHCb collaboration \cite{Aaij:2013mga}. We compare here corresponding theoretical predictions based on both, the first (only $c \to D$) \cite{Maciula:2013wg} and the second ($c + g \to D$) scenario \cite{Nefedov:2014qea},
keeping the same set of $\alpha_{S}$, scales, unPDFs and other details. This comparison is crucial for drawing definite conclusions from double $D$ meson production. As shown in Fig.~\ref{fig:pTsingle}, both prescriptions give a very good description of the LHCb experimental data. Some small differences between them can be observed for both very small and large meson transverse momenta. The latter effect can be relatd to the DGLAP evolution which makes the slope
of the transverse momentum distribution in the second scenario a bit steeper. In the region of very small $p_t$'s the second scenario gives larger cross sections and
slightly overestimates the experimental data points. This may come from the $g \to D$ fragmentation component which approaches a problematic region when $p_t \sim 2 m_{c}$.
Then the treatment of charm quarks as massless in the DGLAP evolution of fragmentation function for very small evolution scale can be a bit questionable and may lead to a small overestimation of the integrated cross sections.

%-----------------------------------------------------------------------------
%------------------------------------------------------------------------------

\begin{figure}[!h]
\begin{minipage}{0.47\textwidth}
 \centerline{\includegraphics[width=1.0\textwidth]{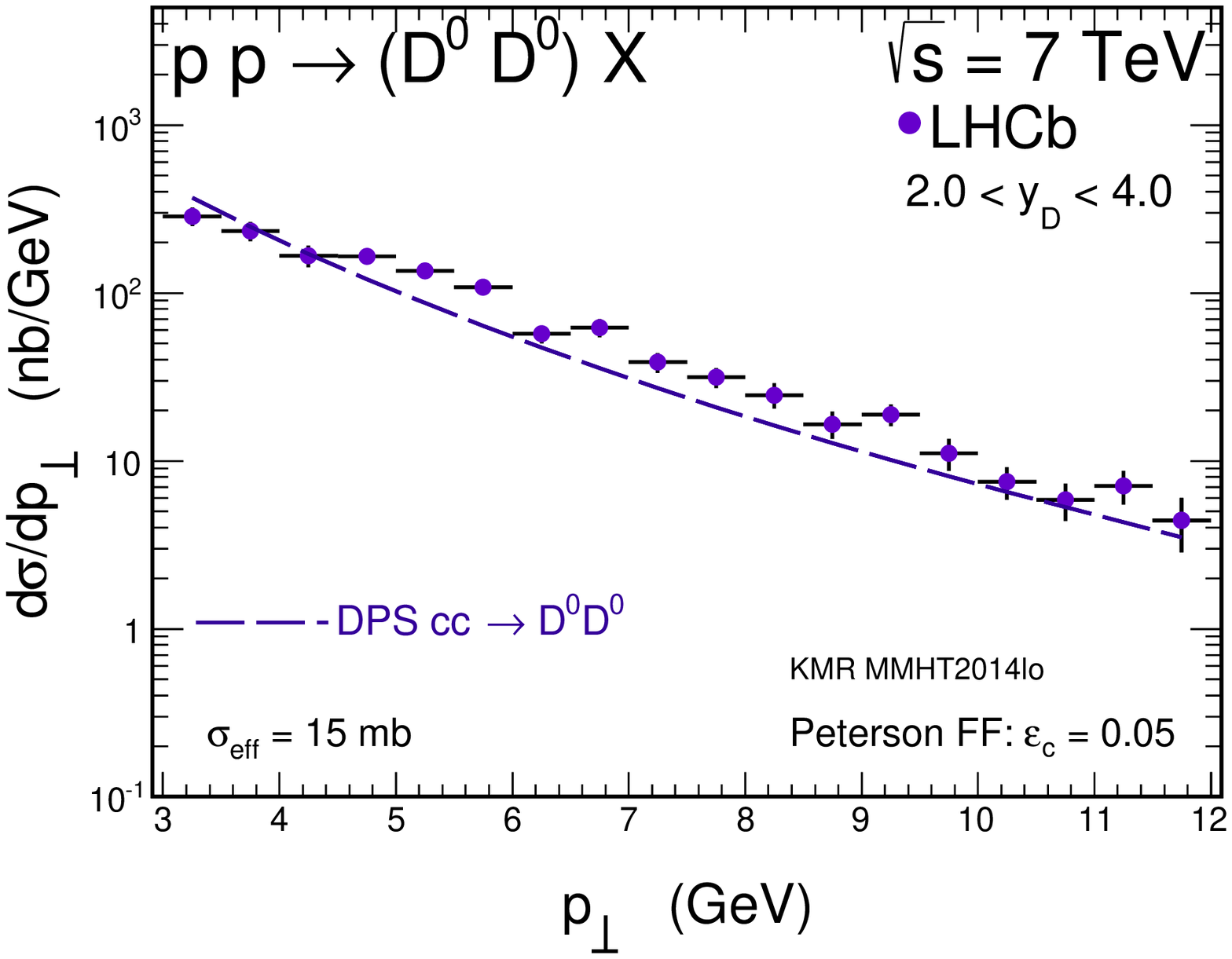}}
\end{minipage}
\hspace{0.5cm}
\begin{minipage}{0.47\textwidth}
 \centerline{\includegraphics[width=1.0\textwidth]{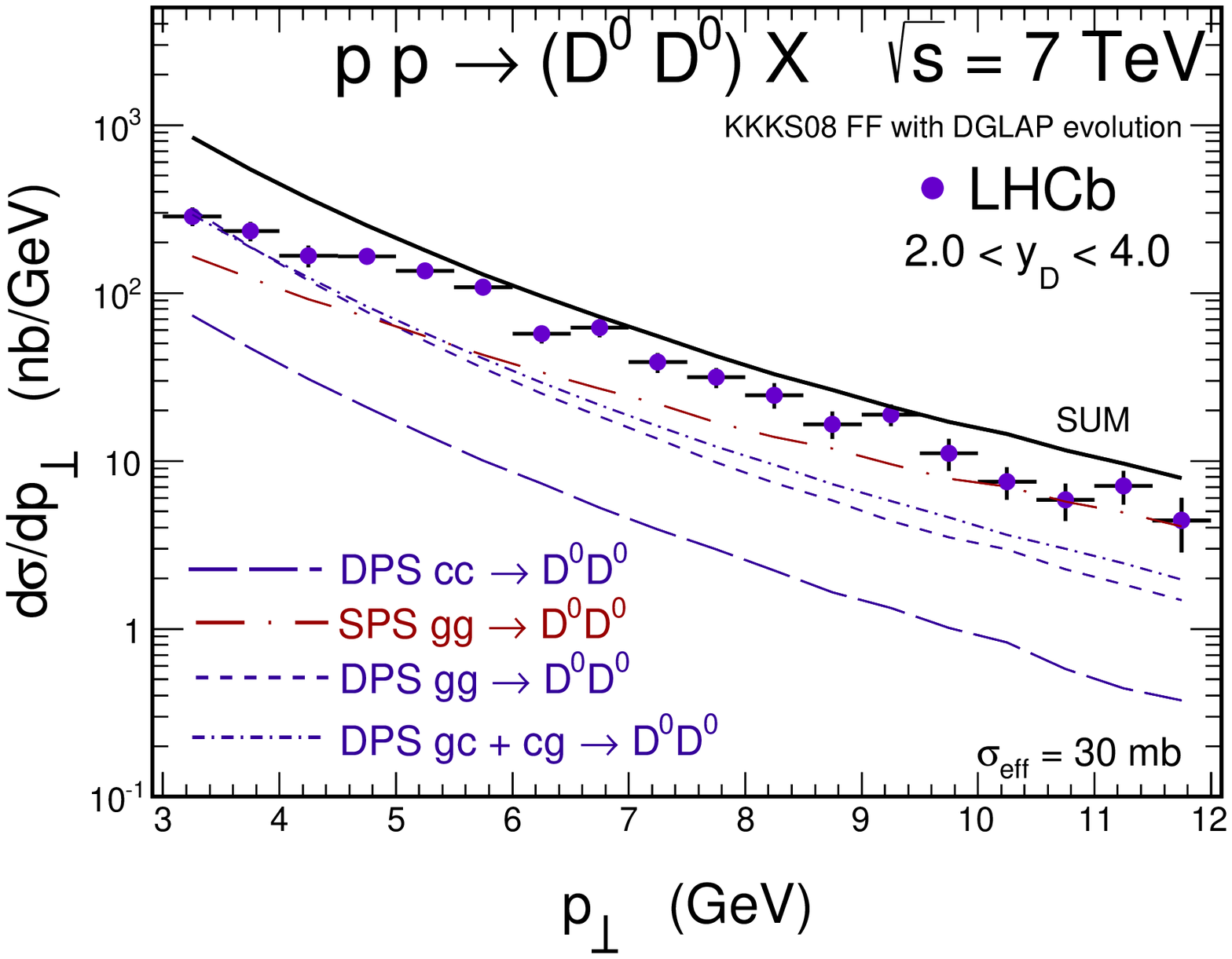}}
\end{minipage}
   \caption{
\small $D^0$ meson transverse momentum distribution within the LHCb
acceptance region.
The left panel is for the first scenario and for the Peterson $c \to D$ fragmentation function
while the right panel is for the second scenario and for the fragmentation function
that undergo DGLAP evolution equation.
 }
 \label{fig:pT}
\end{figure}

Now we go to double charm meson $D^{0}D^{0}$ production.
In Fig.~\ref{fig:pT} we compare results of our calculation
with experimental distribution in transverse momentum of one of the
meson from the $D^{0}D^{0}$ (or $\bar{D}^{0} \bar{D}^{0}$) pair. We show results for the first scenario when standard Peterson FF is used for the $c \to D^0$
(or $\bar c \to {\bar D}^0$) fragmentation (left panel) as well as
the result for the second scenario when the KKKS08 FFs with DGLAP evolution for $c \to D^0$ (or $\bar c \to \bar{D}^0$) and $g \to D^0$ (or $g \to \bar{D}^0$) are used.
One can observe that the DPS $cc \to D^{0}D^{0}$ contribution in the new scenario is
much smaller than in the old scenario. In addition, the slope of the
distribution in transverse momentum changes. Both the effects are due to evolution of
corresponding fragmentation functions. The different new mechanisms give contributions of similar size. We can obtain an agreement in the second case provided $\sigma_{eff}$ parameter is increased from conventional $15$ mb to $30$ mb. Even then we overestimate the LHCb data for $3 < p_{T} < 5$ GeV.
Possible effects that may result in larger value of $\sigma_{eff}$ and in its transverse momentum dependence are discussed in our original paper \cite{Maciula:2016wci}.

%-----------------------------------------------------------------------------
%------------------------------------------------------------------------------
%-----------------------------------------------------------------------------
\begin{figure}[!h]
\begin{minipage}{0.47\textwidth}
 \centerline{\includegraphics[width=1.0\textwidth]{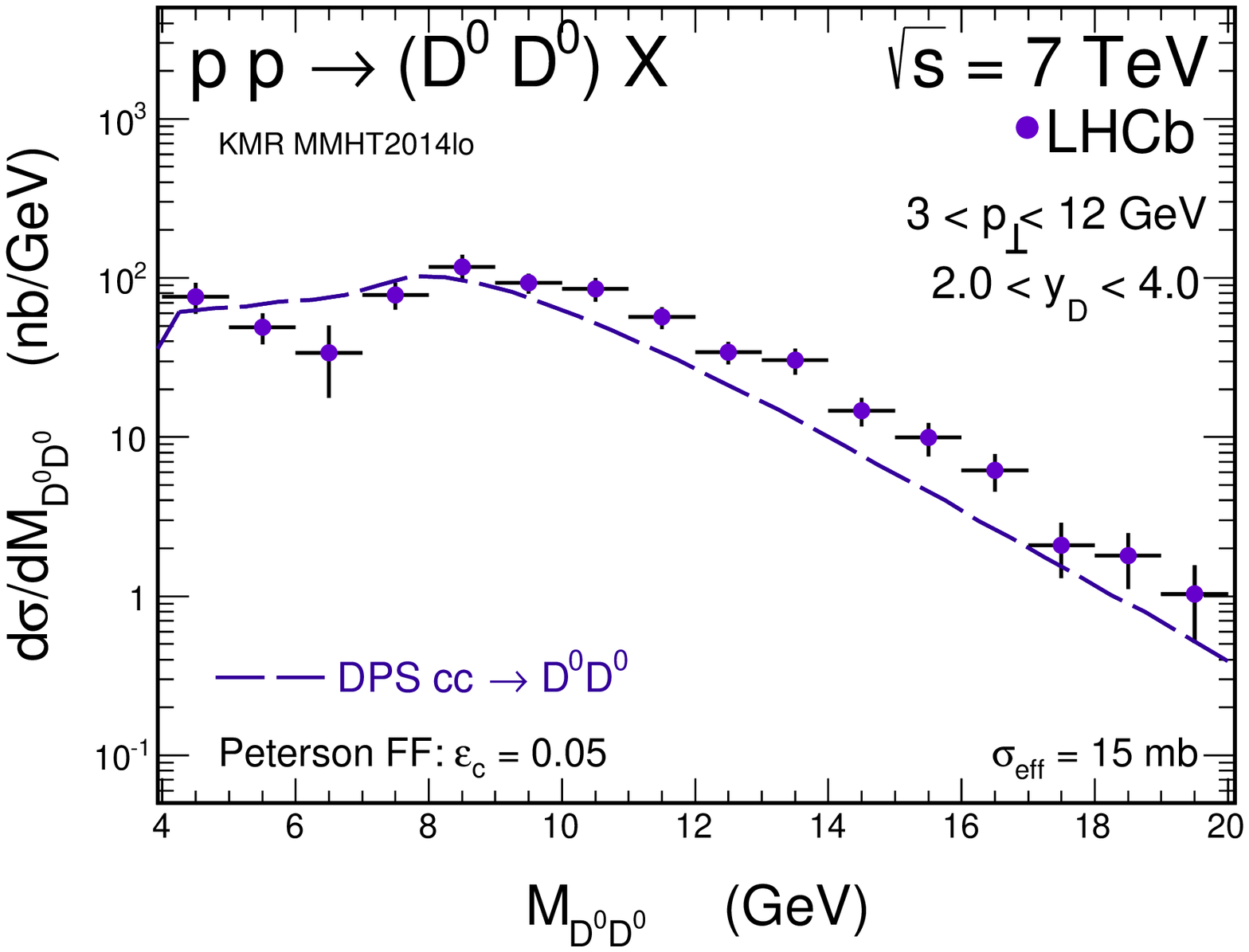}}
\end{minipage}
\hspace{0.5cm}
\begin{minipage}{0.47\textwidth}
 \centerline{\includegraphics[width=1.0\textwidth]{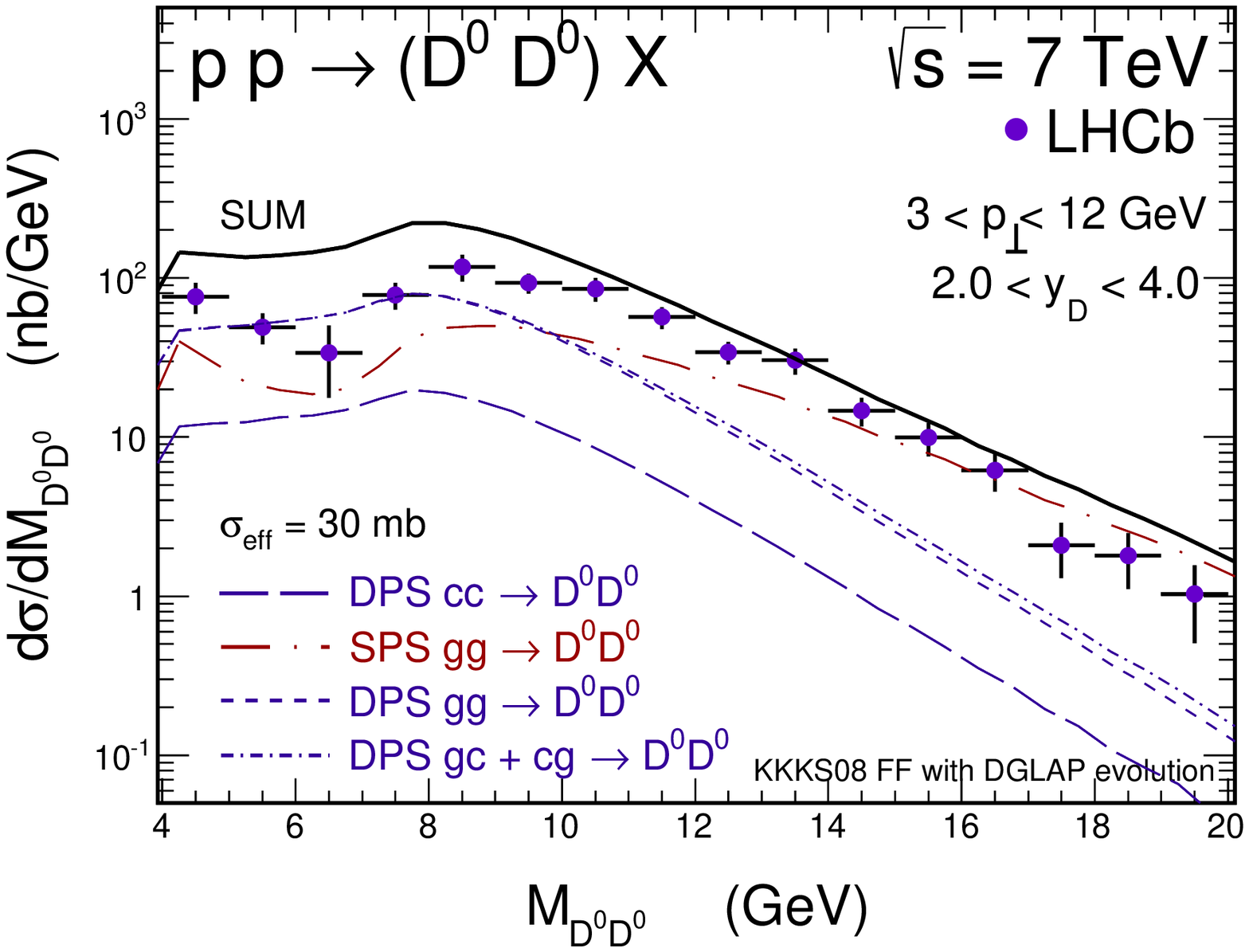}}
\end{minipage}
   \caption{
\small $M_{D^0 D^0}$ dimeson invariant mass distribution within the LHCb acceptance
region.
The left panel is for the first scenario and for the Peterson $c \to D$ fragmentation function
while the right panel is for the second scenario and for the fragmentation function
that undergo DGLAP evolution equation.
 }
 \label{fig:Minv}
\end{figure}
%------------------------------------------------------------------------------
%-----------------------------------------------------------------------------
\begin{figure}[!h]
\begin{minipage}{0.47\textwidth}
 \centerline{\includegraphics[width=1.0\textwidth]{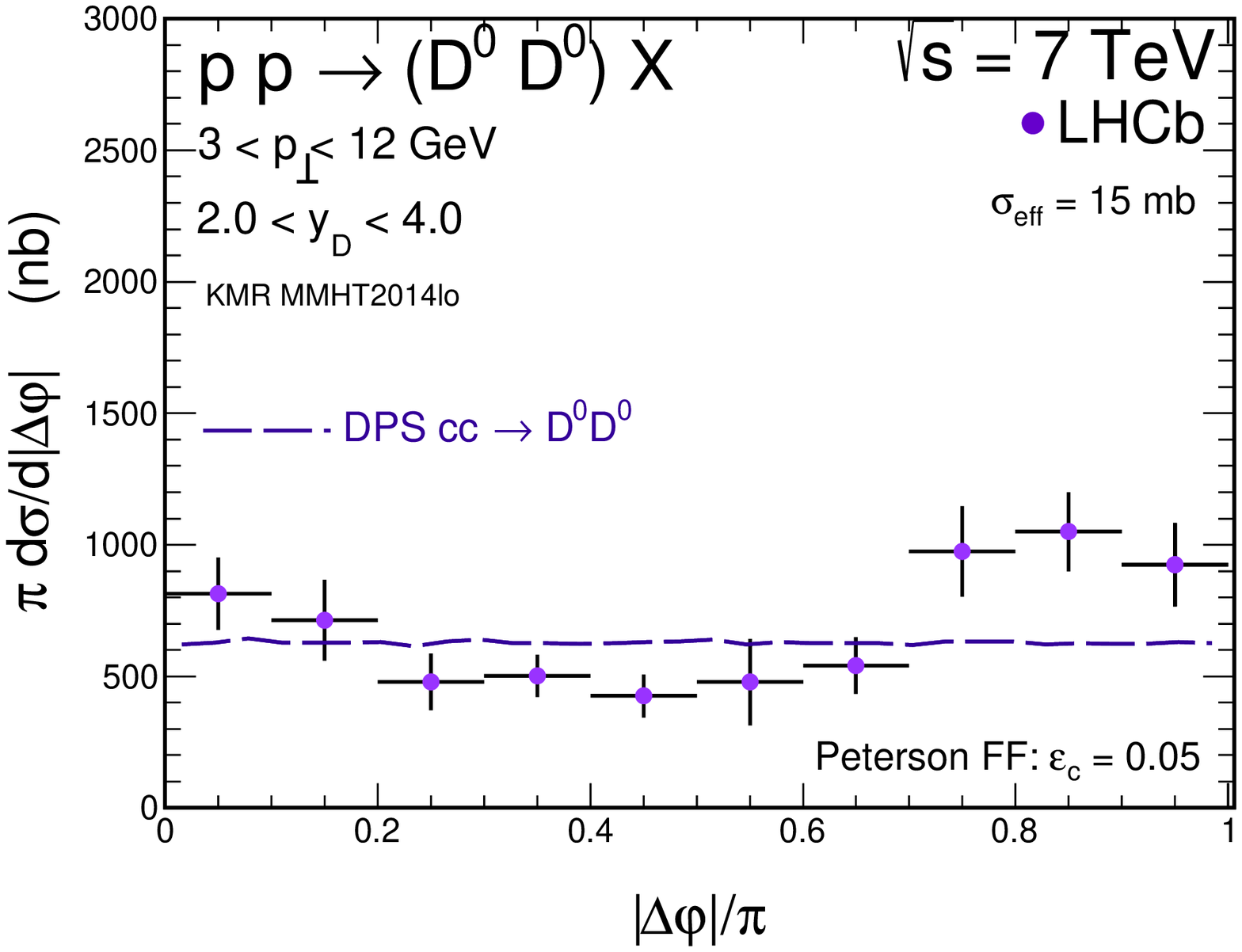}}
\end{minipage}
\hspace{0.5cm}
\begin{minipage}{0.47\textwidth}
 \centerline{\includegraphics[width=1.0\textwidth]{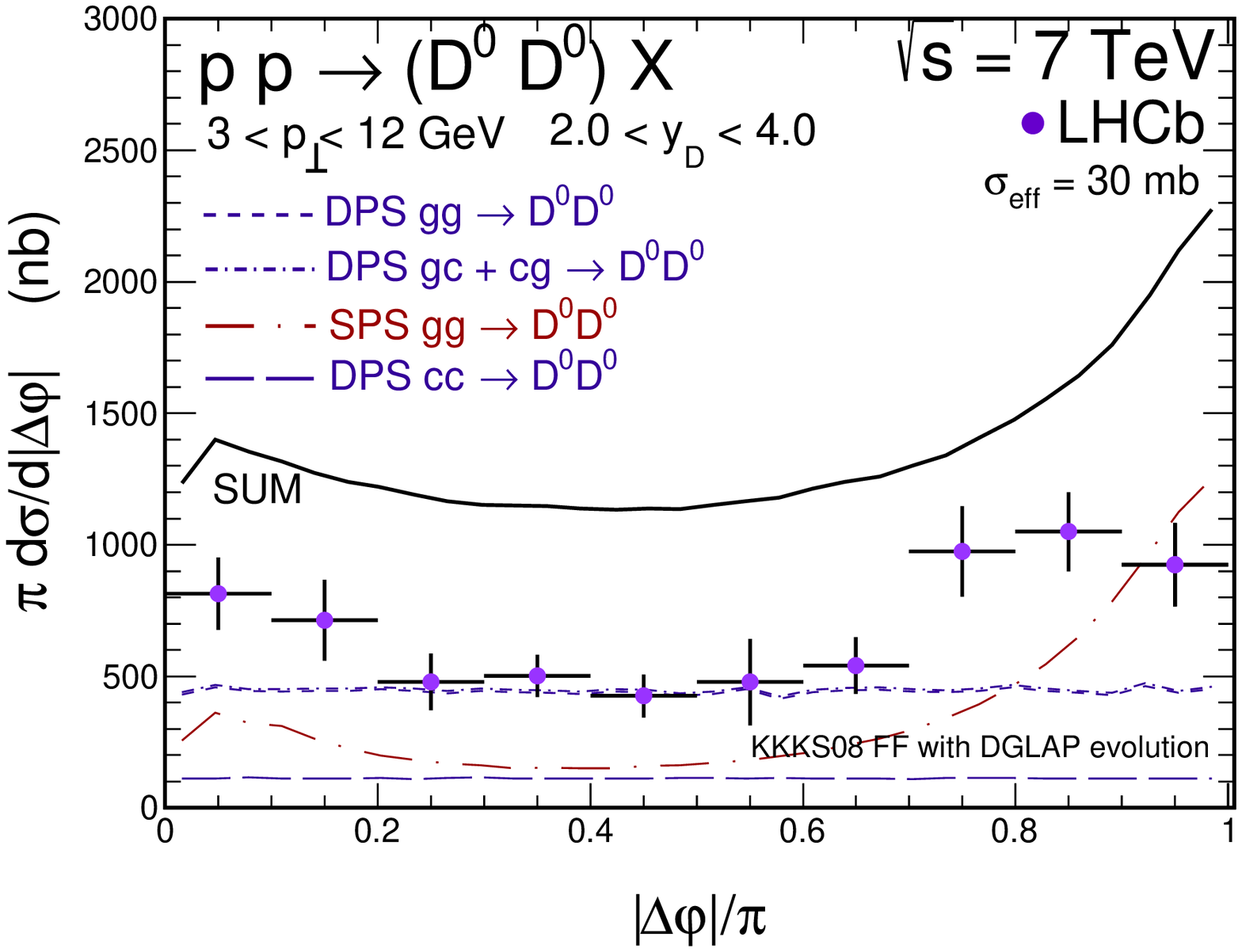}}
\end{minipage}
   \caption{
\small Distribution in azimuthal angle $\varphi_{D^0 D^0}$ between the two $D^0$ mesons
within the LHCb acceptance region.
The left panel is for the first scenario and for the Peterson $c \to D$ fragmentation function
while the right panel is for the second scenario and for the fragmentation function
that undergo DGLAP evolution equation.
 }
 \label{fig:phid}
\end{figure}
%------------------------------------------------------------------------------

In Fig.~\ref{fig:Minv} we show dimeson invariant mass distribution $M_{D^0 D^0}$
again for the two cases considered. In the first scenario we get a good agreement only for small invariant 
masses while in the second scenario we get a good agreement
only for large invariant masses. The large invariant masses are
strongly correlated with large transverse momenta, so the situation 
here (for the invariant mass distribution) is quite similar as 
in Fig.~\ref{fig:pT} for the transverse momentum distribution.

In Fig.~\ref{fig:phid} we show azimuthal angle correlation $\varphi_{D^0 D^0}$ 
between $D^0$ and $D^0$ (or ${\bar D}^0$ and ${\bar D}^0$ mesons).
While the correlation function in the first scenario is completely flat,
the correlation function in the second scenario shows some tendency similar as in the experimental data.

\begin{figure}[!h]
\centering
\begin{minipage}{0.47\textwidth}
 \centerline{\includegraphics[width=1.0\textwidth]{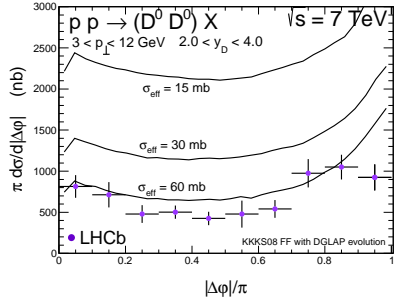}}
\end{minipage}
   \caption{
\small The dependence of the results of the second scenario on the parameter $\sigma_{eff}$ used in the calculation of the DPS contributions.
Here the three lines correspond to $\sigma_{eff}$ equal to $15$, $30$, and $60$ mb, from top to bottom, respectively.}
 \label{fig:sig_eff}
\end{figure}

To summarize the present situation for the second scenario, 
in Fig.~\ref{fig:sig_eff} we show again the azimuthal angle distribution discussed above for different values of $\sigma_{eff}$.
Good description can be obtained only for extremely large values of $\sigma_{eff}$ which
goes far beyond the geometrical picture \cite{Gaunt:2014rua} and that are much larger than for other reactions
and in this sense is inconsistent with the factorized Ansatz. We think that the solution
of the inconsistency is not only in the DPS sector as already discussed in this paper.

%-----------------------------------------------------------------------------
%------------------------------------------------------------------------------

\section{Conclusions}
\label{Summary}

The new scenario with scale-dependent FFs for double $D$ meson production give similar result as the first scenario
with one fragmentation subprocess ($cc \to DD$) and fixed (scale-independent) FFs.
However, correlation observables, such as dimeson invariant mass
or azimuthal correlations between $D$ mesons, are slightly better
described in the second scenario as long as we consider only their shapes. However, to get the proper normalization of the cross sections calculated within the second scenario a much larger value of $\sigma_{eff}$ is needed.

The observed overestimation of the correlation observables in the second scenario comes from the region of small transverse momenta.
It may be related to the fact that the fragmentation function used in the new scenario
were obtained in the DGLAP formalism with massless $c$ quarks and 
$\bar c$ antiquarks which may be a too severe approximation,
especially for low factorization scales (i.e. low transverse momenta)
for fragmentation functions. On the other hand, 
the situation can be also improved when a proper transverse momentum dependence of $\sigma_{eff}$
and/or when perturbative-parton-splitting mechanisms will be included, but this needs further studies.

\vskip+5mm
{\bf Acknowledgments}\\
\vskip+2mm
We thank V.~A. Saleev and A.~V. Shipilova for collaboration in obtaining results presented here.
This study was partially
supported by the Polish National Science Center grant
DEC-2014/15/B/ST2/02528.

\end{document}